\input amstex
\documentstyle{amsppt}

\NoBlackBoxes
\TagsOnRight
\CenteredTagsOnSplits

\magnification=1200
\hcorrection{-0.09 in}
\vcorrection{0.0 in}
\pagewidth{5.6 in}
\pageheight{7.60 in}
\nopagenumbers

\address
L. Chayes
\hfill\newline
Department of Mathematics
\hfill\newline
University of California
\hfill\newline
Los Angeles, California 90095-1555
\endaddress
\email
lchayes\@math.ucla.edu
\endemail

\address
K. Shtengel
\hfill\newline
Department of Physics,
\hfill\newline
University of California,
\hfill\newline
Los Angeles, CA 90095-1547
\endaddress
\email
shtengel\@physics.ucla.edu
\endemail

\topmatter

\title
Critical Behavior for 2d Uniform and \\
Disordered Ferromagnets at Self--Dual Points
\endtitle
\bigskip

\leftheadtext { L. Chayes, K. Shtengel}
\rightheadtext\nofrills {2d Disordered Ferromagnets}

\author
\hbox{\hsize=2.70in
\vtop{\centerline{L. Chayes}
\centerline{{\it Department of Mathematics}}
\centerline{{\ \it University of California, Los Angeles}}
\centerline{{ }}
\centerline{{ }}
\centerline{{ }}}
\vtop{\centerline{K. Shtengel}
\centerline{{\it Department of Physics}}
\centerline{{\ \ \it University of California, Los Angeles}}}}
\endauthor

\thanks
Work supported by the NSA under grant \# MDA904-98-1-0518
\endthanks
\abstract
We consider certain two-dimensional systems with self--dual
points including uniform and disordered $q$--state Potts
models.  For systems with continuous energy density (such as
the disordered versions) it is established that the self--dual
point exhibits critical behavior:  Infinite susceptibility,
vanishing magnetization and power law bounds for the decay of
correlations.
\endabstract

\endtopmatter

\document

\baselineskip = 14.5pt

\vfill

\newpage
\heading
Introduction
\endheading

In this note we will consider the Potts
ferromagnets and related systems on the square lattice.  The Potts models are
defined by the Hamiltonian
$$
\Cal H = -\sum_{\langle x,y \rangle}
J_{x,y}\delta_{\sigma_x,\sigma_y}
\tag 1
$$
with $\sigma_x = 1, 2, \dots, q$ and
$\delta_{\sigma_x,\sigma_y}$ the usual Kronecker delta.  Here
$J_{x,y}$ is non--zero only when $x$ and $y$ are nearest neighbors
and it is assumed that these couplings cannot be negative.

Interest
in the disordered version of these systems has recently been revived,
in particular by J\. Cardy and coworkers \cite{Ca} who have discovered
an apparent close connection between these problems and systems with
random fields.  For the disordered Potts models (among other systems) it was
established in \cite{AW$_1$}, \cite{AW$_2$} that for all temperatures, the
energy
is continuous.  Thus, by conventional definitions, the magnetic ordering
transition is continuous.  However, as pointed out by Cardy in
\cite{Ca} -- as well as in a number of public forums -- for these
systems it has not been established that all aspects of the
transition meet the conventional criteria of a continuous transition:
Vanishing of the order parameter, power law decay for correlations
and infinite susceptibility.  Here we establish that, at least at the
self--dual points, these systems behave critically in the sense of
all the above mentioned (with a power law {\it bound} for decay of
correlations). Our results apply to a variety of systems under the
hypothesis of a continuous energy density.

The method is to employ graphical representations (e\.g\. the random
cluster representation for the Potts models) and in fact applies to
the non--integer cases provided the model is attractive ($q \geq 1$).
In essence, the results here are complimentary to one recently proved
in a paper coauthored by one of us \cite{BC}.  There it was shown
that if (at some point) the energy density is {\it not} continuous
then the discontinuity (a) is unique, (b) occurs precisely at the
self--dual point and (c) coincides with the magnetic ordering
transition.  For these cases the picture is, by and large, complete.
Unfortunately, for the continuous cases, our methods do not rule out the
possibility that the self--dual point is simply a critical point in the interior
of a critical phase.  (Nor does it rule out the possibility that at the
low--temperature edge of this purported phase, the magnetization exhibits a jump
akin to the Thouless effect in one--dimensional long--range systems [T], [AY],
[ACCN].) Nevertheless, taken together, the two sets of results imply that the
self--dual point is always a point of non--analyticity.

This work will be organized along the following lines:  We will start
with the uniform $q \geq 1$ random cluster cases which are the
simplest illustration of the basic method.  Next we will treat
certain straightforward generalizations, e\.g\. the Ashkin--Teller
model and in the second section we treat the disordered Potts model.
\heading
Uniform Systems
\endheading

\subheading{The random cluster model}  We shall begin by
setting notation.  Consider the random cluster model on some
finite connected
$\Lambda
\subset \Bbb Z^{2}$.  If $\omega$ is a configuration of bonds, the
probability of
$\omega$, in the setup with ``free'' boundary conditions is given by
$$
\mu^{q,R}_{\Lambda^f}(\omega)
\propto R^{N(\omega)}q^{C^f(\omega)}
\tag 2
$$
Where $N(\omega)$ is the number of ``occupied'' bonds of the
configuration and $C^f(\omega)$ is the number of connected
components.  If $q$ is an integer, this is the representation of the of
the model described by Equ.(1) -- for free boundary conditions -- with
$J_{i,j} \equiv 1$ and $R = e^\beta -1$.

For other boundary conditions, the formula for the weights must be
modified.  Of primary interest are the {\it wired} boundary conditions
which, back in the spin--system correspond to setting each spin on the
boundary to the same value.  Then the formula for the weights of
configurations is the same as in Equ.(2) with $C^f(\omega)$ replaced by
$C^w(\omega)$ where the latter counts all sites connected to the
boundary as part of the same cluster.

In this note we will restrict attention to the random cluster measures
that are (weak, possibly subsequential) limits of random cluster
measures defined in finite volume.  Here the boundary conditions that we
will consider -- essentially those that are handed down from the
spin--systems -- are defined as follows:  In volume $\Lambda$, the
boundary, $\partial \Lambda$ is divided into $k$ disjoint sets.  All
sites of the individual sets are identified as the same site.  Thus the
boundary consists of $k$ effective sites (components) $v_1, \dots v_k$.
No interior connections between $v_1, \dots v_k$ are permitted.  (I.e\.
any configuration with such a connection is assigned zero weight.)  All
interior sites connected to the same boundary component are considered
as part of the same connected component.  Finally, we will allow
couplings between boundary sites and their neighboring interior sites to
take arbitrary values in $[0,\infty)$. And, of course, we will also
consider arbitrary superpositions of all of the above.

For the case of free and wired boundary conditions, infinite volume
limits, ergodicity etc. follow in a straightforward fashion from the
monotonicity (FKG) properties of the $q \geq 1$ random cluster measures.
(In the disordered cases, some of these points must be rediscussed and
we will do so at the appropriate time.)  We will assume general
familiarity on the part of the reader concerning these properties.
Most of the relevant material can be found in
\cite{ACCN} or
\cite{BC}.  However, if available at the time of reading, the authors
highly recommend the forthcoming article
\cite{GHM}.

The dual model, defined on the lattice $\Lambda^*$ that is dual to
$\Lambda$, has weights of the same form as those in Equ\.(2) with $R$
replaced by $R^* = qR^{-1}$.  The general problem of boundary
conditions for the dual model are a little intricate but for the
purposes of this work, it is sufficient to note that the free and
wired boundary conditions are  exchanged under duality.

In these models (with integer $q$) the relationship between the
bond density in the random cluster model and energy density in the
spin--system is straightforward.  In particular, let $\goth e_{x,y}$
denote the event of an occupied bond that connects the neighboring pair
$\langle x,y\rangle$.  Then, as shown in \cite{CM$_{\text{I}}$},
$$
\langle \delta_{\sigma_x,\sigma_y} \rangle_{\Lambda^\#}^{q,R} =
\frac{1+R}{R}\mu^{q,R}_{\Lambda^\#}(\goth e_{x,y})
\tag 3
$$
where $\langle - \rangle_{\Lambda^\#}^{q,R}$ denotes thermal expectation
in the spin--system in boundary condition $\# = f, w$ (or, for that
matter any other boundary condition).  Thus, in these cases, continuity
in the energy density is manifested as continuity in the bond density.
Hereafter, we will focus on $q \geq 1$ random cluster models and use
continuous {\it bond} density for our working hypotheses.

Let us finally remark that for almost every $R$, the bond density is, in
fact, a well defined concept.  This point (which is fairly standard) has
recently been detailed in \cite{BC} so here we will be succinct.
Consider the free energy, $\Phi(R)$, defined here by
$$
\Phi(R) = \lim_{\Lambda \nearrow \Bbb Z^2}
\frac1{|\Lambda|} \log \Cal Z_{\Lambda^\#}
\tag 4
$$
where $\Cal Z_{\Lambda^\#}$ is the sum of the weights in Equ.(2) and
$\Lambda \nearrow \Bbb Z^2$ means a {\it thermodynamic} limit -- a
regular sequence of boxes.  The function $\Phi(R)$ is convex (as a
function of $\log R$) and hence has a left and right derivative for
every $R$ which agree for almost every $R$.  At points of continuity of
the derivative, there is uniqueness among the translation invariant states
and the bond density in this state is given by the derivative of $\Phi$
(with respect to $\log R$).  At points of discontinuity, the upper value
for the bond density is achieved in the wired state and the lower value in
the free state.

We are ready for our first result:
\proclaim{Theorem 1}
Consider the 2d random cluster models with parameter $q \geq 1$.  Then if
the self--dual point, $R = R^* = \sqrt q$ is a point of continuity of
the bond density, the percolation density vanishes.
\endproclaim
\demo{Proof}
This result can in fact be obtained as a consequence of Theorem 2.1 in
\cite{BC}.  For completeness we will provide a direct proof.  Here we
will establish the contrapositive statement; i.e\. assume that the
percolation probability {\it is} positive at the self--dual point and
show that this implies a discontinuity in the bond density.

Percolation is defined in reference to the wired measures (and limits
thereof).  These measures are ergodic under $\Bbb Z^2$ translations,
respect the $x,y$--axis symmetry and have the FKG property.  In short,
these measures satisfy all the conditions of the theorem in \cite{GKR}
which forbids coexisting infinite clusters of the opposite type.  Thus,
with probability one, whenever there is percolation, all dual bonds
reside in finite clusters.  However, if there is percolation (in the
wired state) at the self--dual point, the same cannot be said in the
limiting free boundary condition measure.  Indeed, from the perspective
of the {\it dual} bonds, this is a wired state.  Hence, in this state, the dual
bonds percolate and the regular bonds do not.  It is thus evident that the
limiting free and wired measures are distinct;
$\mu^{q,R}_{\Lambda^f}(-)$ is strictly below $\mu^{q,R}_{\Lambda^w}(-)$.
By the (corollary to) Strassen's theorem (see \cite{L} p. 75) this
implies that the bond density in the wired state is strictly larger than
that of the free state; the self--dual point is thus not a point of
continuity for the derivative of $\Phi$
\qed
\enddemo
\proclaim{Corollary}  Under the hypotheses of Theorem 1, there is a unique
limiting Gibbs state/ random cluster measure at the self--dual point.
\endproclaim
\demo{Proof}
It was proved in \cite{ACCN}
(Theorem A.2) that the absence of percolation (for $q \geq 1$) implies that a
unique limiting random cluster measure and, for (integer $q \geq 2$) a unique
Gibbs state in the corresponding spin--system.  So this follows immediately.
\qed
\enddemo
Let $A$ and $B$ denote disjoint sets in $\Bbb Z^2$.  We will denote by
$\{A \longleftrightarrow B\}$ the event that some site in $A$ is
connected to some site in $B$ by occupied bonds.  Further, if $D$
contains both $A$ and $B$, we let
$\{A \longleftrightarrow B\}_{D}$ denote the event that such a
connection occurs by a path that lies entirely in $D$.
For $x$ and $y$ (distinct) points in $\Bbb Z^2$, let $g_{x,y} =
g^{\#}_{x,y}(R,q) = \mu^{q,R}_{\#}(\{x \longleftrightarrow y\})$
be the probability that $x$ and $y$ belong to the
same cluster.  We will call this object the {\it connectivity
function}.  For integer
$q = 2, 3,
\dots$ the connectivity function is equal to (or proportional to) the
spin--spin correlation function.  In these cases, the susceptibility and
the average cluster size are also identified.  For non--integer $q$, the
geometric quantities are {\it defined} to be the objects of interest.
These quantities are the subject of our next theorem which is a direct
consequence of Theorem 1.  The proof below borrows heavily from the
argument in \cite{A}.
\proclaim{Theorem 2}
For self--dual $q \geq 1$ random cluster models with vanishing
percolation probability, the function $g_{x,y}$ has a power law lower
bound.  Explicitly, if $0$ is the origin and $\bold{L}$ is the point
$(L,0)$ on the $x$ axis then $g_{0,\bold{L}} \geq \frac 18
L^{-2}$.  Finally, the average cluster size is infinite.
\endproclaim
\demo{Proof}
Since there is a unique limiting state then, in particular, the limiting free
and wired measures coincide.
Consider the $L\times L$ square centered at the origin which we denote by
$S_L$.  In every configuration, there is either a left--right crossing by
regular bonds or a top--bottom crossing by dual bonds.  By duality, in
the limiting measure, these probabilities are both one half.
\footnote{To ensure that this is strictly true, one must carefully
construct the square so that it is {\it exactly} self--dual.  However,
for the arguments here, it is actually sufficient to observe that one of
the probabilities must be greater or equal to one half.}
Letting $\Cal L_L$ denote the left edge and $\Cal R_L$ the right edge of
the square, this implies that
$$
\sum_{x \in \Cal L_L, y \in \Cal R_L} g_{x,y} \geq \frac 12.
\tag 5
$$
Hence, for some (deterministic) $x^* \in \Cal L_L$ and $y^* \in \Cal R_L$
we have
$$
g_{x^*,y^*} \geq \frac 1{2L^2}.
\tag 6
$$
This is, in essence the bound on the correlation function.  For \ae
sthetic purposes we will show that a similar bound holds for
$g_{0,\bold{L}}$ but let us first attend to the susceptibility.
Following the logic of Eqs.(5) and (6), there is an $x^{**}$ in
$\Cal L_L$ that is connected to $\Cal R_L$ by a path inside $S_L$
with probability of order $L^{-1}$:
$$
\mu^{q, R = \sqrt q}(\{x^{**} \longleftrightarrow \Cal R_L\}_{S_L})
\geq \frac 1{2L}.
\tag 7
$$
Regarding the point $x^{**}$ as being at the center of a square of side
$2L$ and using translation invariance, we find
$$
\mu^{q, R = \sqrt q}(\{0 \longleftrightarrow \partial S_{2L}\})
\geq \frac 1{2L}
\tag 8
$$
i.e\. $X_L \equiv \mu^{q, R = \sqrt q}(\{0 \longleftrightarrow \partial
S_{L}\}) \geq L^{-1}$.  This immediately implies a divergent
susceptibility/cluster size.  Indeed writing
$$
\Cal X = \sum_x g_{0,x} = \sum_{L}\sum_{x\in\partial S_L}g_{0,x}
\geq \sum_{L} X_L
\tag 9
$$
this result follows.
Finally, let us obtain our bound for the correlation function along the
coordinate axes.  Consider the event
$(\{x^{**} \longleftrightarrow \Cal R_L\}_{S_L}$ along with along with
its mirror image reflected along the midline of the square.  I.e\. a
connection between $\Cal L_L$ and $y^{**}$ by a path inside $S_L$ where
$y^{**} = x^{**} + \bold{L}$.  If these two connections occur in tandem
with a top bottom crossing of $S_L$, we achieve a connection between
$y^{**}$ and $x^{**}$.  These events are all positively correlated hence
$$
g_{0,\bold{L}} = g_{y^{**}, x^{**}} \geq
(\frac 12)(\frac1{2L})(\frac1{2L}).
\tag 10
$$
\qed
\enddemo

It is clear that the above generalizes to other random cluster systems.
However since at present there are not too many examples of physically
relevant models that satisfy all of the required conditions, we will be
content with a small selection.
\subheading{The [r,s]-cubic (generalized Ashkin--Teller) model}
Consider two copies of $\Bbb Z^2$ with two sets of Potts spins:
$\tau_i \in \{1, \dots r\}$ and $\kappa_i \in \{1, \dots s\}$.  It is
convenient to envision the model as two layers of $\Bbb Z^2$, the
$\tau$--layer and the $\kappa$--layer with the $\tau$--layer just above
the $k$--layer.  In any case, the Hamiltonian is given by
$$
H = -\sum_{\langle x,y \rangle}
[a\delta_{\tau_i,\tau_j}\delta_{\kappa_i,\kappa_j}
+ b\delta_{\kappa_i,\kappa_j} +
c\delta_{\tau_i,\tau_j}]
\tag 11
$$
where, as it turns out, we will be interested in the cases $a,b,c \geq 0$.

The dual relations for this model (at least for integer $r$ and $s$)
were derived some time ago in \cite{dN}, \cite{DR} by algebraic methods.
(Of course the special case $r = s = 2$ was derived much earlier
starting, in fact, with \cite{AT}.)  More recently, graphical
representations for this model have been discovered
\cite{CM$_{\text{I}}$} \cite {PfV}, (and see also \cite{SS}) in which the
duality is manifest.  Consider bond configurations
$\omega = (\omega_\tau, \omega_\kappa)$ i.e\. separate bond
configurations in the $\tau$-- and $\kappa$--layers. As usual, we will
start in finite volume.  Let $N(\omega_\tau)$ and $N(\omega_\kappa)$
denote the number of occupied bonds in the $\tau$-- and $\kappa$--
layers respectively.  Let $N(\omega_\tau \vee \omega_\kappa)$ denote
the number of edges where at least one of the $\tau$-- or $\kappa$--
layers have occupied bonds and finally let $N(\omega_\tau \wedge
\omega_\kappa)$ denote the number of edges where both the $\tau$-- and
$\kappa$-- layers have occupied bonds.  The graphical representation is
defined by the weights
$$
W(\omega) = \bold{A}^{N(\omega_\tau \vee \omega_\kappa)}
\bold{B}^{N(\omega_\tau \wedge \omega_\kappa)}
\bold{C}^{[N(\omega_\kappa) - N(\omega_\tau)]}
r^{C(\omega_\tau)}s^{C(\omega_\kappa)}
\tag 12
$$
where $C(\omega_\tau)$ and $C(\omega_\kappa)$ are the number of
connected components as in the usual random cluster problems. (And
typically must be augmented with some boundary conditions.)
The relationship between $\bold{A}$, $\bold{B}$ and $\bold{C}$ and
$a$, $b$ and $c$ is as follows:
$$
\bold{A} = [(e^{\beta b} -1)(e^{\beta c} -1)]^{\frac 12}
\tag 13a
$$
$$
\bold{B} = \frac{e^{\beta (a + b + c)} - e^{\beta b} - e^{\beta c} + 1}
{[(e^{\beta b} -1)(e^{\beta c} -1)]^{\frac 12}}
\tag 13b
$$
$$
\bold{C} = \left[\frac{(e^{\beta b} -1)}
{(e^{\beta c} -1)}\right]^{\frac 12}.
\tag 13c
$$
In order for the graphical representation to make sense, we require $b,
c \geq 0$.  However, this is not the case with $a$ but it turns out
that the FKG property -- which we will need -- only holds if $\bold{B}
\geq \bold{A}$ \cite{BC} thus we actually require all couplings in
Equ\.(13) to be ferromagnetic.  Under these conditions for the case $r
= s$, $b = c$ it was shown in \cite{CM$_{\text{I}}$} that there is a
single ordering transition as the temperature is varied.

The dual model is defined straightforwardly:  edges of the dual lattice
in, say, the $\tau$--layer that are traversal to occupied bonds are
considered vacant dual bonds, those edges traversal to vacant bonds are
the occupied dual bonds and similarly in the $\kappa$--layer.
The duality conditions are easily obtained from the weights in
Equ\.(12) (for the simple reason that $\vee \leftrightarrow \wedge$
under duality) and the result is:
$$
\bold{A}^* = \frac{\sqrt {rs}}{\bold{B}}
\tag 14a
$$
$$
\bold{B}^* = \frac{\sqrt {rs}}{\bold{A}}
\tag 14b
$$
$$
\bold{C}^* = \sqrt{\frac rs} \thinspace \frac 1{\bold{C}}
\tag 14c
$$
The analog of Theorems 1 and 2 for this system are readily established:
\proclaim{Theorem 3}
Consider the $(r,s)$--cubic model as described with $\bold{A} \geq
\bold{B}$.  Let $\Phi(\bold{A},\bold{B},\bold{C})$ denote the free energy
similar to that defined in Equ.(4). Suppose that a self dual point:
$(\bold{A}, \bold{B}, \bold{C}) = (\bold{A}^*,\bold{B}^*, \bold{C}^*)$,
is a point of continuity for any first derivative of $\Phi$.  Then the
percolation probability in either layer vanishes.  Let $g^{\tau}_{x,y}$
and $\Cal X^\tau$ denote the connectivity function and average size of
clusters in the $\tau$--layer and similarly for
$g^{\kappa}_{x,y}$ and $\Cal X^\kappa$.  Then $g^{\tau}_{0,\bold{L}} \geq
\frac 18 L^{-2}$, $\Cal X^\tau$ is infinite and similarly for
$g^{\kappa}_{x,y}$ and $\Cal X^\kappa$.
\endproclaim
\demo{Proof}
It is convenient, but not essential, to restrict attention to the
``plane'' $\bold{C} = \bold{C}^*$.  Indeed, following the argument below,
it can be shown that continuity with respect to $\bold{A}$ and
$\bold{B}$ actually implies continuity with respect to $\bold{C}$ thus,
for all intents and purposes, the $\bold{C}$--variable is out of the
play.  Continuity of the derivative with respect to $\bold{B}$ implies that
at the point $(\bold{A}, \bold{B}, \bold{C})$, the density of ``doubly
occupied'' bonds is independent of state for any translation invariant
state.  Add to this continuity of the derivative with respect to
$\bold{A}$ and (since
$N(\omega_\tau \vee \omega_\kappa) + N(\omega_\tau \wedge \omega_\kappa)
= N(\omega_\tau) + N(\omega_\kappa)$) we may conclude that the total bond
density is the same in every translation invariant state.  However, we
claim that this implies the same result for the separate densities.
Indeed if the $\tau$--density were discontinuous, to keep the total
density continuous would require a compensating discontinuity in the
$\kappa$--density.  But these densities are positively correlated; i.e\.
the discontinuities must go in the same direction.  In particular, we
could find a state (at the point $(\bold{A}, \bold{B}, \bold{C})$) where
both densities achieved their lower value and another where they both
obtain the upper value.

Constancy of the bond density implies no percolation in either layer at
a self--dual point which in turn implies unicity of the state.  The rest
of the argument follows mutatis mutandis the proof of Theorem 2.
\qed
\enddemo
\subheading{A loop related model}  Our final example appeared in the
context of loop models in \cite{CPS}.  Let $\omega$ denote a bond
configuration on $\Bbb Z^2$ and let $\tilde \omega$ denote the
{\it complimentary} configuration on the dual lattice:  If a bond of
$\omega$ is occupied than so is the traversal bond and similarly for
vacancies.  (In other words, the vacant bonds of the dual configuration
are the occupied bonds of the complimentary configuration.)  The weights,
in finite volume are given by
$$
V(\omega)  =  L^{N(\omega)}s^{C(\omega)}s^{C(\tilde \omega)}.
\tag 15
$$
We remark that from a technical perspective, unrelated boundary
conditions for $\omega$ and $\tilde \omega$ may be implemented.  However
it is natural to assert that if one is fully wired so is the other and
similarly with free boundary conditions.  A derivation identical to the
one for the usual random cluster model shows that the dual model is the
same model with the parameter
$$
L^* = \frac{s^2}{L}
\tag 16
$$
along with the usual exchange of boundary conditions.  For double--free
or double--wired (as well as other) boundary conditions, the FKG property
follows easily:
\proclaim{Proposition 4}
For free or wired boundary conditions, the random cluster models defined
defined by the weights in Equ.(15) with $s \geq 1$ have the FKG property.
Further, for these boundary conditions, if $R_1 \geq R_2$ and $s_1 \leq
s_2$ the measure with parameters $(R_1,s_1)$ FKG dominates the one with
parameters $(R_2,s_2)$.
\endproclaim
\demo{Proof}
For the usual random cluster model, the FKG property follows from the FKG
lattice condition \cite{FKG}:  Let $\omega_1$ and $\omega_2$ denote two
bond configurations, $\omega_1 \wedge \omega_2$ the configuration
of bonds occupied in both $\omega_1$ and $\omega_2$ and $\omega_1
\vee \omega_2$ the configuration of bonds occupied in either.
The lattice condition reads:
$\mu(\omega_1 \wedge \omega_2)\mu(\omega_1 \vee \omega_2)
\geq \mu(\omega_1)\mu(\omega_2)$ which follows because
$C(\omega_1 \wedge \omega_2) + C(\omega_1 \vee \omega_2)
\geq C(\omega_1) + C(\omega_2)$ \cite{ACCN}.  In the present case, we
need only apply this argument twice, once to $C(\omega)$ and once to
$C(\tilde \omega)$.  The FKG dominance follows by writing the one set of
weights as an increasing function times the other.
\enddemo
\remark{Remark}
It would appear that the model under discussion is very close to the $q =
s^2$--state random cluster model.  This follows by noting that for the
former, $C(\omega)$ and $C(\tilde \omega)$ are identically distributed.
Then, we may write $s^{C(\omega) + C(\tilde \omega)} =
s^{2C(\omega)}s^{C(\tilde \omega) - C(\omega)}$ and suppose that the
``fluctuations'' $C(\tilde \omega) - C(\omega)$ are (thermodynamically)
small.  However, at present there is no hard evidence of such an
equivalence.  On the other hand, the self--dual point can in fact be
realized as the endpoint of the self--dual line of the symmetric (r = s,
C = 1) cubic model corresponding to $A \to \infty$ and $B \to 0$.  Here
the $\tau$--bonds may be taken to be the occupied bonds and the
$\kappa$'s to be the vacants.  The condition $B = 0$ ensures that they
cannot coincide while $A = \infty$ implies one bond or the other actually
is occupied.
\endremark
\proclaim{Theorem 5}  For the model defined by the weights in Equ.(15),
the results of Theorem 3 apply; If $R = s$ is a point of continuity for
the bond density then the connectivity function has a power law lower
bound and the average cluster size is infinite.
\endproclaim
\demo{Proof}
Follows from the same arguments as the proofs of Theorems 1 and 2
\qed
\enddemo
\remark{Remark}
In this model, the results of \cite{BC} also apply:  If there is any
point of discontinuity for the bond density, that point must be the self
dual point.  (The cases of the Potts model and the cubic model for $r =
s$ were the explicit subject of \cite{BC}; the identical arguments
apply to the current case.)  Thus, one way or another in all these systems the
self--dual points are points of ``phase transitions''.  For large $q$, $r$,
and/or $s$ -- at least in the integer cases -- it is straightforward to
show that discontinuities {\it do} occur.  (Theorem IV.2 in
\cite{CM$_{\text{I}}$} covers all of these cases.)  The difficulty is the
opposite cases:  establishing continuity.  Indeed, to the authors'
knowledge, the only non--trivial case where this has been done with
complete rigor is the Ising magnet.  However, the next section features
systems where the required continuity has been guaranteed.  $\Bbb b$
\endremark
\newpage
\heading
Quenched Potts Models
\endheading
For the remainder of this paper, we will deal exclusively with the
$q$--state Potts model as defined by the Hamiltonian in Equ.(1); we will
treat the case where the $J_{x,y}$ are (non--negative) independent random
variables.  (And ultimately to prove theorems along the lines of Theorems
1 and 2, we will need to focus on distributions that are self--dual.)  We
strongly suspect that with only minimal labor the forthcoming
results could be extended to disordered versions of the various other models
discussed in the previous section. But here we will focus on the minimal case.

The approach in this work will be somewhat different from the
usual mathematical studies of disordered systems: rather than looking at
properties that are ``typical'' of configurations of couplings, we will
construct, from the outset, the {\it quenched} measure -- more precisely,
the graphical representation thereof.  When all the preliminaries are in
place, this has the advantage of allowing a derivation that is
essentially indistinguishable from the uniform cases.  The disadvantage
is that many of the ``basic preliminaries'' will require some attention.
\subheading{The quenched measure}
Let $\Lambda \subset \Bbb Z^2$ (or $\Bbb Z^d$ for the duration of the
preliminaries) denote a finite volume.  In what follows, the inverse
temperature $\beta$ as well as the value of $q$ ($\geq 1$) will be regarded as
fixed and hence will be suppressed notationally.  Let
$\eta = \{J_{x,y}\mid \langle x,y \rangle \in \Lambda\}$ denote a set of
couplings.  Let $\#$ denote a boundary condition on $\partial \Lambda$.
In general we will allow the boundary condition to depend on the
realization of couplings so we will write $\#(\eta)$.  (For the case of
continuous variables $J_{x,y}$ we must also stipulate that $\#(\eta)$ is a
{\it measurable} function.)  We let
$\langle - \rangle^\eta_{\Lambda;\#(\eta)}$ denote the finite volume Gibbs
state (for {\it this} realization of couplings and {\it this} boundary
condition.)  Similarly, we may consider the random cluster measures
$\mu^\eta_{\Lambda;\#(\eta)}(-)$.  Our assumption about the
$J_{x,y}$ is that they are i.i.d. non--negative variables.  Let $\goth
b(-)$ denote the product measure for configurations of couplings and
$\Bbb E_{\goth b}(-)$ the expectation with respect to this measure.  Then
the quenched measures are defined as the $\goth b$--averages of the
``thermal'' averages according to
$\langle - \rangle^\eta_{\Lambda;\#(\eta)}$ and
$\mu^\eta_{\Lambda;\#(\eta)}(-)$.  Explicitly, if
$F(\sigma_{x_1}, \dots \sigma_{x_k})$ is a function of spins (with
$x_1, \dots x_k \in \Lambda$) then the quenched average of $F$ is given by
$$
\overline{\langle F \rangle}_{\Lambda;\#} =
\Bbb E_{\goth b}(\langle F \rangle^\eta_{\Lambda;\#(\eta)}).
\tag 17a
$$
Similarly, for a bond event $\Cal A$,
$$
\overline{\mu}_{\Lambda;\#}(\Cal A) =
\Bbb E_{\goth b}(\mu^\eta_{\Lambda;\#(\eta)}(\Cal A)).
\tag 17b
$$
Most of our attention will be focused on the quenched random cluster
measures as defined in Equ.(17b) -- or the infinite volume limits
thereof.  Our first proposition establishes some FKG properties of these
quenched measures:
\proclaim{Proposition 6}
On finite $\Lambda$, let $w$ and $f$ denote the boundary conditions that
are, respectively wired and free for all $\eta$.  Then the measures
$\overline {\mu}_{\Lambda,w}(-)$
and
$\overline {\mu}_{\Lambda,f}(-)$
have the FKG property (in the sense of positive correlations).
Furthermore, for any $\# = \#(\eta)$, the wired measure
$\overline{\mu}_{\Lambda, w}(-)$ FKG--dominates the measure
$\overline{\mu}_{\Lambda, \#}(-)$
\endproclaim
\demo{Proof} For the FKG property, we will just do the wired case; the free case
is identical. Let $\Cal A$ and $\Cal B$ denote increasing events (defined on the
bonds of $\Lambda$).  By the FKG properties of the measures
$\mu^\eta_{\Lambda;w}(-)$,
$$
\mu^\eta_{\Lambda;w}(\Cal A \cap \Cal B) \geq
\mu^\eta_{\Lambda;w}(\Cal A ) \mu^\eta_{\Lambda;w}(\Cal B).
\tag 18
$$
Now we observe that for any increasing event $\Cal C$, the quantity
$\mu^\eta_{\Lambda;w}(\Cal C )$ is an increasing function of $\eta$.
Indeed, if $\eta_1 \succ \eta_2$ (meaning $J_{x,y}^1 \geq J_{x,y}^2$ for
each bond in $\Lambda$) then the random cluster measure with couplings
$\eta_1$ FKG dominates the one with couplings $\eta_2$.  But then
$$
\align
\overline {\mu}_{\Lambda,w}(\Cal A & \cap \Cal B)
= \Bbb E_{\goth b}(\mu^\eta_{\Lambda;w}(\Cal A \cap \Cal B))
\geq \Bbb E_{\goth b}(\mu^\eta_{\Lambda;w}(\Cal A )
\mu^\eta_{\Lambda;w}(\Cal B)) \geq \\
& \Bbb E_{\goth b}(\mu^\eta_{\Lambda;w}(\Cal A ))
\Bbb E_{\goth b}(\mu^\eta_{\Lambda;w}(\Cal B ))
= \overline {\mu}_{\Lambda,w}(\Cal A)
 \overline {\mu}_{\Lambda,w}(\Cal B).
\tag 19
\endalign
$$
The same works for the free case (and various other boundary conditions
that are independent of $\eta$).
Finally, the stated FKG--domination is an obvious consequence of the fact
that for each $\eta$, the measure $\mu^\eta_{\Lambda, w}(-)$ FKG--dominates
the measure $\mu^\eta_{\Lambda, \#(\eta)}(-)$
\qed
\enddemo
As a corollary, we obtain the existence of infinite volume limits as in
the usual random cluster cases:
\proclaim{Corollary}  The infinite volume limits $\overline{\mu}_f(-)$
and $\overline{\mu}_w(-)$ exist in the sense that if
$\Lambda_k \nearrow \Bbb Z^d$ is any sequence of boxes with
$\Lambda_{k+1} \supset \Lambda_k$ then the limits of
$\overline{\mu}_{\Lambda_k,f}(-)$ and $\overline{\mu}_{\Lambda_k,w}(-)$
exist and are independent of the sequence.  These limiting measures are
translation invariant and invariant under exchange of coordinate axes.
\endproclaim
\demo{Proof}
The argument is exactly as in the standard proofs and is a consequence of
the following observation:  If $\Lambda_1 \subset \Lambda_2$ then for any
fixed $\eta$, the restriction of $\mu^\eta_{\Lambda_2,w}(-)$ to
$\Lambda_1$ is FKG dominated by the wired measure in $\Lambda_1$.  Thus
the same statement holds for the quenched average of these measures. A
similar sort of domination, but in the opposite direction is established
for the free measures.  The remainder of the proof is now identical to the
derivations for uniform random cluster models (with occasional use of
translation invariance and coordinate symmetry of $\goth b (-)$).  Such
proofs have been written in many places (see e.g\. \cite{CM$_{\text{I}}$}
Theorem 3.3) and need not be repeated here.
\qed
\enddemo

We now demonstrate that absence of percolation is the correct criterion
for uniqueness.  Our working definition of percolation is fairly standard:
\definition{Definition}
Let $\Lambda \subset \Bbb Z^d$ be a finite set that contains the origin.
We define
$$
P_{\infty} = \lim_{\Lambda \nearrow \Bbb Z^d}
\overline{\mu}_{\Lambda,w}(0 \longleftrightarrow \partial \Lambda).
\tag 20
$$
We say there is percolation if $P_\infty$ is not zero.
\enddefinition
\remark{Remark}
It is obvious, by the considerations of the corollary to Proposition 6
that this limit exists.  Further, if $P_\infty$ vanishes, there is no
percolation by any other criterion.  Finally it is not difficult to show
that $P_\infty$ is exactly the spontaneous magnetization in the
spin--system.
\endremark

Next we establish the quenched analog of Theorem A.2 in \cite{ACCN}.
\proclaim{Proposition 9}
If $P_\infty = 0$ there is a unique limiting quenched random cluster
measure and a unique limiting quenched Gibbs measure.
\endproclaim
\demo{Proof}
Our proof will in essence be to show that any sequence of finite volume
measures converges to the free measure.  Let $\Cal A$ denote any local
increasing event.  Let $\Lambda$ denote a large (finite) box -- the bonds
of which determine the event $\Cal A$.  Now consider a much larger box
$\Xi$ along with some boundary condition $\#(\eta)$; the measure
$\overline{\mu}_{\Xi,\#}(-)$ may be thought of as ``well along the way''
towards the construction of some infinite volume measure.

Since the percolation probability is assumed to vanish, it is clear that
if $\Xi$ is sufficiently large, then, for $\epsilon > 0$
$$
\overline{\mu}_{\Xi,\#}
(\partial \Lambda \longleftrightarrow \partial \Xi)
\leq \epsilon^2.
\tag 21
$$
Thus if $\Cal D_{\Lambda, \Xi} =
\{\eta\mid \mu^\eta_{\Xi,\#(\eta)}(\partial \Lambda \longleftrightarrow
\partial \Xi) > \epsilon\}$ then
$\goth b (\Cal D_{\Lambda, \Xi}) < \epsilon$.

Now for any $\eta \in \Cal D_{\Lambda, \Xi}$, with
$\mu^\eta_{\Xi,\#(\eta)}$--probability greater than $1-\epsilon$, there
is a ``ring'' (separating surface) of vacant bonds in the region between
$\partial \Lambda$ and $\partial \Xi$.  Conditioning to the ``outermost''
such ring gives us a measure which, in the interior of the ring, is
equivalent to free boundary conditions on the ring.  For
any $\eta$, this in turn is dominated by the measure with free boundary
conditions on
$\partial \Xi$ and dominates (in $\Lambda$) the measure with free boundary
conditions on $\partial \Lambda$.  Thus, for
$\eta \in \Cal D_{\Lambda, \Xi}$,
$$
(1-\epsilon)[\mu^\eta_{\Lambda, f}(\Cal A)]
\leq \mu^\eta_{\Xi, \#(\eta)}(\Cal A) \leq
(1-\epsilon)[\mu^\eta_{\Xi, f}(\Cal A)] + \epsilon
\tag 22
$$
and hence
$$
(1-\epsilon)[\overline{\mu}_{\Lambda, f}(\Cal A)]
\leq \overline{\mu}_{\Xi, \#}(\Cal A) \leq
(1-\epsilon)[\overline{\mu}_{\Xi, f}(\Cal A)] + 2\epsilon
\tag 23
$$
where the extra $\epsilon$ comes from the
$\epsilon \notin \Cal D_{\Lambda, \Xi}$.
From Equ.(23) it is easy to see that all sequences of finite volume
quenched measures converge to the limiting free measure.

The argument for the uniqueness of the quenched Gibbs state follows from
the above by noting that the thermal average of any local spin--function
can be expressed as expectations of random cluster functions (which
themselves are finite combinations of increasing events.)  This proves
(a) the existence of a limiting $\overline{\langle - \rangle}_{f}$ (and
for that matter a limiting $\overline{\langle - \rangle}_{w}$) and (b)
that if the magnetization vanishes that this is the unique limiting state.
\qed
\enddemo
The final result we will need is the ergodic property for the free and
wired quenched random cluster measures.
\proclaim{Theorem 10}
The measures $\overline{\mu}_w(-)$ and $\overline{\mu}_f(-)$ are ergodic
under $\Bbb Z^d$ translations.
\endproclaim
\demo{Proof}
We will do the wired case, the free case is nearly identical.
Let $\Cal A$ and $\Cal B$ denote local events assumed, without loss of
generality, to be increasing.  Let $\vec r \in \Bbb Z^d$ and let
$T_{\vec r}(\Cal B)$ denote the event $\Cal B$ translated by $\vec r$.
We will show that
$\lim_{\vec r \to \infty} \overline{\mu}_w(\Cal A \cap T_{\vec r}(\Cal B))
= \overline{\mu}_w(\Cal A)\overline{\mu}_w(\Cal B)$.

By FKG and translation invariance we have, for any $\vec r$,
$$
\overline{\mu}_w(\Cal A \cap T_{\vec r}(\Cal B))
\geq \overline{\mu}_w(\Cal A)\overline{\mu}_w(\Cal B).
\tag 24
$$
Now consider $|\vec r|$ large -- far larger than the scale of the regions
that determine the events $\Cal A$ and $\Cal B$.  Let $s \leq |\vec r|$
be chosen so that $\Lambda_s$, the box of side $s$ centered at the origin
and its translate by $\vec r$, which we denote by
$T_{\vec r}(\Lambda_s)$, are disjoint but within a few lattice spacings
of each other.  Finally, let us consider an $L$ that is very large
compared with $\vec r$; we will approximate
$\overline{\mu}_w(\Cal A \cap T_{\vec r}(\Cal B))$ by
$\overline{\mu}_{\Lambda_L, w}(\Cal A \cap T_{\vec r}(\Cal B))$.  By the
FKG property,
$\overline{\mu}_{\Lambda_L, w}(\Cal A \cap T_{\vec r}(\Cal B))$ is less
than the corresponding probability given that all bonds on the outside of
$\Lambda_s$ and $T_{\vec r}(\Lambda_s)$ are occupied.  But given these
occupations, the measure inside $\Lambda_s$ is equivalent to wired
boundary conditions on $\Lambda_s$ and similarly for
$T_{\vec r}(\Lambda_s)$.  Now for each $\eta$, the wirings make these
interior measures independent.  Thus we have
$$
\mu^\eta_{\Lambda_L, w}(\Cal A \cap T_{\vec r}(\Cal B)) \leq
\mu^\eta_{\Lambda_s, w}(\Cal A)
\mu^\eta_{T_{\vec r}(\Lambda_s), w}(T_{\vec r}(\Cal B)).
\tag 25
$$
However as functions of $\eta$, the two objects on the right of Equ.(25)
are independent -- they take place on disjoint sets.  It is clear that
$\mu^\eta_{T_{\vec r}(\Lambda_s), w}(T_{\vec r}(\Cal B))$ averages to
$\overline{\mu}_{\Lambda_s, w}(\Cal B)$ and thus
$$
\overline{\mu}_{\Lambda_L, w}(\Cal A \cap T_{\vec r}(\Cal B)) \leq
\overline{\mu}_{\Lambda_s, w}(\Cal A)
\overline{\mu}_{\Lambda_s, w}(\Cal B).
\tag 26
$$
Letting $L \to \infty$ we get
$$
\overline{\mu}_{w}(\Cal A \cap T_{\vec r}(\Cal B)) \leq
\overline{\mu}_{\Lambda_s, w}(\Cal A)
\overline{\mu}_{\Lambda_s, w}(\Cal B).
\tag 27
$$
and hence
$$
\lim_{\vec r \to \infty}\overline{\mu}_{w}(\Cal A \cap T_{\vec r}(\Cal B))
\leq \lim_{s \to \infty}
\overline{\mu}_{\Lambda_s, w}(\Cal A)
\overline{\mu}_{\Lambda_s, w}(\Cal B)
= \overline{\mu}_{w}(\Cal A) \overline{\mu}_{w}(\Cal B).
\tag 28
$$
This completes the proof for the wired case, the free case works the same
way, here we use decreasing events for $\Cal A$ and $\Cal B$.
\qed
\enddemo
\subheading{Main results}
We are ready for the disordered analogs of Theorems 1 and 2.  However, in
this case, we will not need to hypothesize the required continuity:
this is the central subject of \cite{AW$_1$}, \cite{AW$_2$}.  Let us first
briefly discuss duality in the disordered case.  In the general setup, let
$$
J^*(J) = \log\left[  1 + \frac q{e^J - 1} \right].
\tag 29
$$
(I.e\. $e^J - 1 = q/[e^{J^*} - 1]$).
Then what is needed, in the discrete case, to have $\beta = 1$ a point of
self--duality is that $\goth b(J_{x,y} = J) = \goth b(J_{x,y} =
J^*(J))$.  (A similar statement holds for continuous or other
distributions)  Indeed, if this is the case we see that the probability of
bonds and dual bonds of equivalent {\it strength} are the same.  Then, in
finite volume, the probabilities of two coupling configurations that are
equivalent under duality (including the usual exchange of boundary
conditions) are equal.

Sometimes it is convenient to parameterise the distribution and
allow
$\beta$ to vary.  For example, suppose there are two bond values $J_1$ and
$J_2$ with
$\goth b(J_{x,y} = J_1) = \goth b(J_{x,y} = J_2) = 1/2$.  Since
temperature is back in the problem, we may assume, without loss of
generality that $J_1 = 1$ and write $J_2 = \lambda$ with
$0 \leq \lambda \leq 1$.  Then, in the $\lambda, \beta$ plane it is not
hard to see that the model with parameters $\lambda, \beta$ is equivalent
under duality to the one with parameters $\lambda^*, \beta^*$ where
$$
\beta^* = \log\left[ 1 + \frac{q}{e^{\lambda \beta} -1} \right]
\tag 30a
$$
and
$$
\lambda^* = \frac{\log\left[ 1 + \frac{q}{e^{\beta} -1} \right]}
{\log\left[ 1 + \frac{q}{e^{\lambda \beta} -1} \right]}.
\tag 30b
$$
The system is self--dual $(\lambda = \lambda^*, \beta = \beta^*)$ when
$(e^{\beta} -1)(e^{\lambda \beta} -1) = q$.

\proclaim{Theorem 1$^\prime$}
Consider a disordered Potts model of the type described at a self--dual
point.  Then there is a unique limiting state with zero magnetization.
\endproclaim
\demo{Proof}
The proof is the same as the proof of Theorem 1 which we will
recapitulate for continuity.  If the magnetization (percolation
probability) were non--zero then in the free boundary state, the
percolation probability for dual bonds would be non--vanishing.
Proposition 6 and Theorem 10 allow us to use the result in \cite{GKR}.
Thus the free and wired states would be distinguished.  But these are FKG
ordered states so it would follows that the bond (and hence energy)
density would differ in the two states implying a discontinuity in the
energy density (or bond density).  This however, is contradicted by the
results of \cite{AW$_1$}, \cite{AW$_2$}.  Proposition 9 connects the absence of
magnetization to uniqueness.
\qed
\enddemo
\proclaim{Theorem 2$^\prime$}
Under the hypotheses of Theorem 1$^\prime$, in the limiting state the
quenched correlation function satisfies
$$
\overline{\langle \delta_{\sigma_0,\sigma_L} - \frac 1q \rangle}
\geq \frac 18 \frac 1{L^2}.
$$
Further the quenched susceptibility, defined as
$$
\overline{\Cal X} =
\sum_y\overline{\langle \delta_{\sigma_0,\sigma_y} - \frac 1q \rangle}
$$
is infinite.
\endproclaim
\demo{Proof}
Again this follows exactly the proof of Theorem 2 once we can conclude
that the probability of a square crossing is one half.  Although this
follows from self--duality on ``general principles'' it is comforting to
consider the square $L \times L$ square, $S_L$ in the middle of a finite
(but much larger) square with wired boundary conditions on the top and
right and with free boundary conditions on the left and bottom.  Then the
quenched crossing probability is manifestly exactly one half and the
thermodynamic limit can be taken, which gets us to our unique state, and we
conclude that the probability in the limiting state is one half.
\qed
\enddemo
\bigskip
\baselineskip = 10pt

{\eightpoint L.C. takes great pleasure in thanking M. Aizenman; in general
for many useful conversations over the years and in particular for some
tips on this problem. And for the last reprint of \cite{AW$_2$}.}

\enddocument
\end